# Modelling Polymerization-Induced Self Assembly (PISA)

Ruslan Shupanov[1], Pavel Kos[1], Alexei Gavrilov[1], Alexander Chertovich[2,1]

[1] – Lomonosov Moscow State University, Physics Department.
[2] – Semenov Institute of Chemical Physics

**Abstract**

In this work we studied polymerization-induced self-assembly by means of computer simulations. Using this model, phase diagrams of the micelle states were constructed depending on the polymer concentration and the asymmetry of the composition for various reaction conditions. We found that if the reaction is ideal controlled radical polymerization (the initiation speed is much larger than the propagation speed and there are no side reactions such as termination or chain transfer), the phase diagram is no different from that obtained for pre-synthesized monodisperse diblock-copolymers with one insoluble block. Next, we studied two cases of slow initiation. We found that the phase diagram change dramatically: upon decreasing the initiation speed, the regions of spherical and cylindrical micelles shrink, while the region of vesicles/lamellae expands. This happens because at small initiation speed there is a significant amount of chains with a very short non-soluble block (and even without such block altogether), which do not participate in the formation of micelles. Therefore, decreasing the initiation speed essentially "remaps" the phase diagram coordinates by making the effective concentration lower (by decreasing the number of active chains) and the effective block length ratio higher (again, because the number of active chains gets higher, while the number of monomers remains the same).

## 1.Introduction

Structured block copolymer systems have attracted huge attention due to a large number of application areas, including targeted drug delivery systems, nanoreactors, emulsion stabilizers (for micelles of diblock copolymers) [1], nanolithography, and a variety of nanostructured membranes, including those electrochemical applications and for the filtration of liquids (for microphase-separated copolymers). However, when using classical diblock copolymers, a complex multi-step procedure (including multi-stage synthesis) is required to prepare the system, which also imposes limitations on certain system parameters (for example, the concentration of the solution in the case of micelles [1]). To overcome these difficulties, the method called polymerization induced self-assembly (PISA) can be used.

The PISA method is very young - the first articles began to appear about 10 years ago [2–4]. The standard approach in this method is to grow the second block on a pre-synthesized homopolymer; the growing block is solvophobic and tends to precipitate, but the first block, which is solvophilic, stabilizes the micelles. A great advantage of PISA in comparison with the classical method of obtaining micelles from diblock-copolymers prepared in advance is that it is possible to use substantially higher polymer concentrations [1]. Two classes of systems can be distinguished: emulsion PISA [2,4,5] and dispersive PISA [6–8]. In the former case, the monomer is initially insoluble in the selected solvent, so the polymerization takes place in the droplets enriched with it, whereas in the latter case, the incompatibility between the solvent and the solvophobic monomer is not so large, and only its homopolymers of a certain length are insoluble. In this project, we will be interested in the second variant (dispersion PISA), since it allows us to control the result of polymerization and structuring much more precisely. It should be noted that most articles that investigate PISA use the reversible addition-fragmentation chain transfer (RAFT) polymerization (for example, [6]), but there are examples of the use of atom-transfer radical polymerization (ATRP) [3] and nitroxide-mediated polymerization (NMP) [4]. We can mention the work [1] as a fresh and concise review of the recent advances in PISA. It indicates that most researchers obtain spherical micelles during PISA, but in the work [6] phase diagram in coordinates (the length of the insoluble block-the concentration of the polymer) was constructed; classical structures were found: spherical micelles, cylindrical micelles and vesicles. Thus, PISA allows one to control the type of the resulting structures.

As for the study of PISA in theory and simulation, there is no comprehensive research at the time. In the recent paper [9], phase diagram of micelles formed by diblock copolymers obtained by homopolymerization starting from a soluble macroinitiator was obtained using the dissipative particle dynamics method. Depending on the length of the macroinitiator, the authors obtained various structures including spherical micelles, cylindrical micelles, vesicles, rings and layers. However, only one solution concentration was investigated in [9], the lengths of the resulting copolymers were extremely small (the maximum average length was 12), and the stability of the initial dispersion of monomers was not investigated and raises doubts - a very large incompatibility between the solvent and the monomers of growing blocks was used.

This work is aimed at developing a robust model for simulation of PISA as well as constructing phase diagrams of micellar states for a wide range of system parameters. We are also interested in studying the influence of the reaction conditions to achieve better correspondence with the existing experimental data.

## 2. Method and model

### 2.1 Method

Simulations were carried out using the dissipative particle dynamics (DPD) method [2,3]. This method is a version of the coarse-grained molecular dynamics; the polymer chains are represented beads-and-springs model. The logic of coarse-grained methods implies that every bead is not generally speaking a model of a single atom, but rather of atomic groups like a monomer unit or a statistical segment of a polymer chain. The internal degrees of freedom of such objects are neglected, which leads to a significant acceleration of the calculations. The beads are moving in a continual space, and their motion obeys the Newton's laws. Generally speaking, the DPD method allows one to study phenomena on time and length scales significantly larger than those available for atomistic simulations. The reason for that is not only the aforementioned reduction in the number of simulated objects, but also the usage of "soft" potentials, which do not diverge as the distance between two beads approaches 0. Usually, a linearly decaying force that vanishes at distances larger than some specific value (so called cutoff radius) is used[3] (so the potential is quadratic). While such potentials seem to be chosen arbitrary at the first glance, have a rather simple underlying physical sense: if one averages the potential between the centers of mass of two atomic groups (for example, monomer units) over time using full-atomic molecular dynamics, the potential very similar to that used in DPD will be the result [4,5]; therefore, one obtains some kind of a "time coarse-graining". The soft potential also allows one to use larger timesteps for integrating the equations of motion compared to the molecular dynamics. Another interesting feature of DPD is that normally simulated chains are phantom (that is, their connections can pass "through each other"), which does not affect the phase behavior, but allows reaching the equilibrium state significantly faster. In the work [3] a direct relationship between the parameters of DPD method and the parameter of the classical Flory-Huggins theory was established. In general, the DPD method is now widely used to simulate polymer systems with different polymer concentrations, ranging from single molecules to melts [6,7]. Detailed description of the simulation methodology could be found elsewhere [3].

In all our calculations we used the following parameters: DPD number density $\rho=3$; integration timestep $\Delta t=0.04$; bond length $l=0$; bond stiffness $K=4.0$; DPD conservative parameter between alike particles $a_{ii}=25.0$. Flory–Huggins parameter $\chi$ was calculated using common expression $\chi_{ij} \approx 0.3\Delta a_{ij}$ from the work [3].

**2.2 Model**

In our model of the PISA process, the system initially contains 3 components:1) pre-synthesized linear precursors of the fixed length $n_A=4$; 2) monomers which form the block B during the subsequent reaction; 3) solvent S. The precursor beads are solvophilic ($\chi_{AS}=0$), while the B-monomers as well as B-monomer units are solvophobic ($\chi_{BS}=1.5$); for simplicity, the incompatibility between the A and B beads were set to $\chi_{AB}=1.5$ as well. Each precursor serves as an initiator for the radical polymerization reaction, during which diblock-copolymers are formed. The $\chi_{BS}$-value is not high enough to cause precipitation of the B-monomers, but B-blocks start to aggregate when they are long enough. In our study, the phase diagrams were calculated in the coordinates ($n_B/n_A$ - total polymer concentration), where $n_B/n_A$ is averaged over all the resulting chains in the system and represents the resulting diblock-copolymer asymmetry, and the total polymer concentration is taken at the maximum conversion when all (or almost all) the B-monomer is spent.

**2.3 Reaction scheme**

In order to simulate the radical polymerization reaction, we used the well-known Monte-Carlo approach;[8–11] this method has recently been used to simulate radical copolymerization in bulk[12] and pores[13] as well as emulsion polymerization,[14] and the results were found to be in qualitative and even quantitative agreement with the experimental data, which proves that this approach is a good choice for studying PISA. In short, the main idea is rather simple: every $\tau_0$ timesteps the active chain ends can attach monomers that are located closer than $R_{chem}$ to them with the probability of $p_p$ (propagation probability); the attached monomers become the active ends themselves. In this work we did not consider the side reactions such as termination or chain transfer (they will be studied in the next work); therefore, the reaction is characterized by only 2 parameters: the propagation probability $p_p$ and the initiation probability $p_i$. The reaction radius $R_{chem}$ was chosen to be equal to 1.0, i.e. to the interaction potential cutoff distance $R_c$.

We set the time interval between reaction steps $\tau_0 = 200$ DPD steps. This value is large enough to have local spatial equilibration in the nearest surrounding of each active center. At the same time, it is small enough to simulate a nearly continuous (non-discrete) process and to obtain high total monomer conversions in a reasonable computational time.

# 3. Results and discussions

## 3.1 Micelle formation in solution of pre-synthesized diblock-copolymers

In order to have a reference system for PISA, we started with investigating the case of pre-synthesized diblock-copolymers; this is a classical system which has been studied in a number of theoretical and experimental works[1,15]. Diblock-copolymers were placed into the simulation box at $\chi=0.5$ so that initially the system is homogeneous, and then the $\chi$-value was gradually increased with a step of $\Delta\chi=0.05$ up to $\chi=1.5$ (the systems were relaxed for $10^6$ steps at each intermediate $\chi$-value); this mimics a realistic experiment where there is a finite speed of the temperature change. The length of the insoluble block B ($n_B$) and the total polymer concentration ($c_p$) were varied. The length of the soluble block A($n_A$) was fixed at 4. Fig. 1 shows the obtained phase diagram.

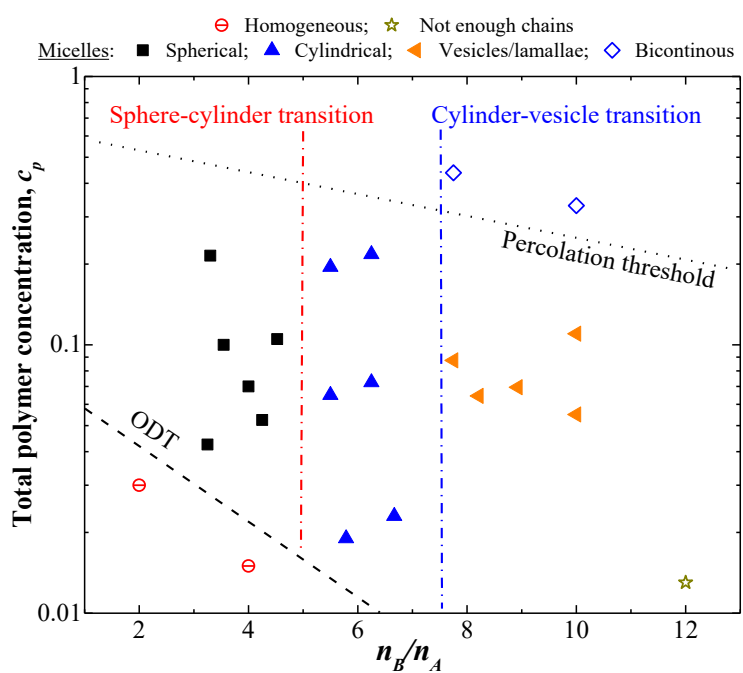

Fig.1 The phase diagram of the micelle formation of pre-synthesized diblock copolymers in solution. The transition lines show the expected behavior and are depicted as a guide for the eye.

It contains the classical phases: spherical and cylindrical micelles as well as vesicles; their snapshots are presented in Fig.2. The hollow markers represent percolating structures which are observed at high polymer concentrations. In general, the position of the phases on the diagram corresponds to the theoretical predictions;[15] the percolating structures will be discussed later.

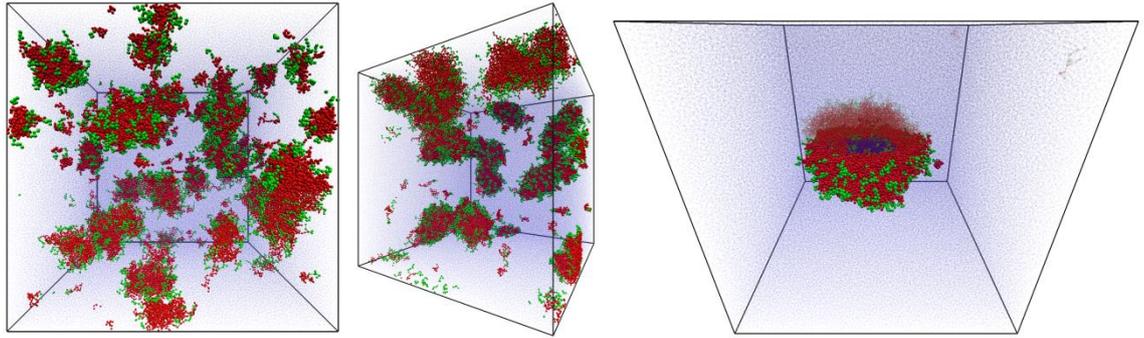

Fig.2 Typical snapshots of the obtained micelles: spherical (left, $n_B/n_A=4.52$, $c_p=5$); cylindrical (middle $n_B/n_A=5.5$, $c_p=6.5$); vesicles ($n_B/n_A=10$, $c_p=5.5$, right). The insoluble blocks are depicted in red, the soluble blocks are green, while the solvent is blue. The latter is rendered semi-transparent for better visibility. To obtain the former two structures, $n_A$ was taken to be equal to 6 for better segregation.

The structure observed at $c_p=1.2$ and $n_B/n_A=12$ (marked as "not enough chains") is located in the area where vesicles should be observed, but due to the small number of chains in the systems only a single spherical micelle is formed in the simulation box.

**3.2 Micelle formation in solution of during PISA with ideal reaction**

Let us now study the case of PISA. We started with the simplest case of the initiation probability $p_i=1$, which correspond to the case of immediate initiation, which, given that the propagation probability $p_p=0.04$ is much smaller, leads to the situation when all the chains start to grow almost simultaneously. The obtained phase diagram is presented in Fig.3.

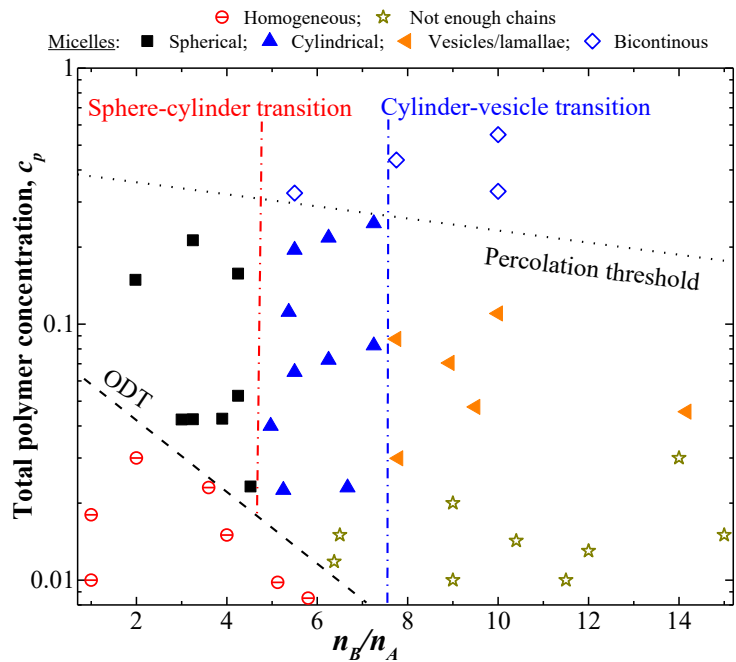

Fig.3 The phase diagram obtained for the "ideal" PISA process with $p_i=1$. The transition lines show the expected behavior and are depicted as a guide for the eye.

Surprisingly, this phase diagram is no different from that obtained in the case of monodisperse pre-synthesized diblock-copolymers; the only significant difference is an increased number of "free" chains (not forming a micelle) in the solution. In order to understand the reason of this behavior, let us study the obtained copolymers. Fig. 4 depicts typical B-block length distribution as well as their dispersity Đ obtained for three main structures of micelles (sphere, cylinder and vesicle).

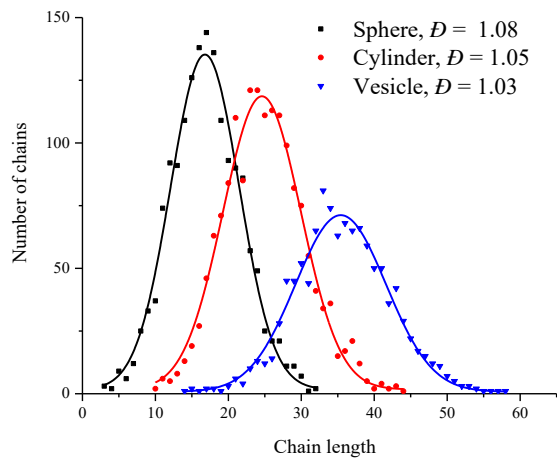

Fig.4 The insoluble block length distributions obtained for different morphologies: spherical micelles ($c_p$=5.25 and $n_b/n_a$=4.25), cylindrical micelles ($c_p$=7.25 and $n_b/n_a$=6.25) and vesicle ($c_p$=6.951 and $n_b/n_a$=8.93).

We can clearly see that the distributions are very narrow. The largest dispersity of Đ=1.08 is observed for the case of spherical micelles, which is obviously due to the fact that spheres are formed at the shortest overall chain length in our simulations, since the soluble precursor length is fixed. In our simulations the chains are phantom (i.e. they can pass through each other), which greatly improves the equilibration dynamics reduces the probability of the formation of kinetically trapped states. Given all that, the fact that the diagrams coincide is expected.

**3.3 Micelle formation in solution of during PISA with slow initiation**

In reality, however, the reaction is not ideal: there are side reactions such as termination and chain transfer as well as slow initiation. The effect of the side reactions is a topic of a separate study and will be studied in our next work; here we will focus on the effect of the initiation speed. In order to do that, we calculated two additional phase diagrams at $p_i$=0.1 and $p_i$=0.01; they are presented in Fig.5.

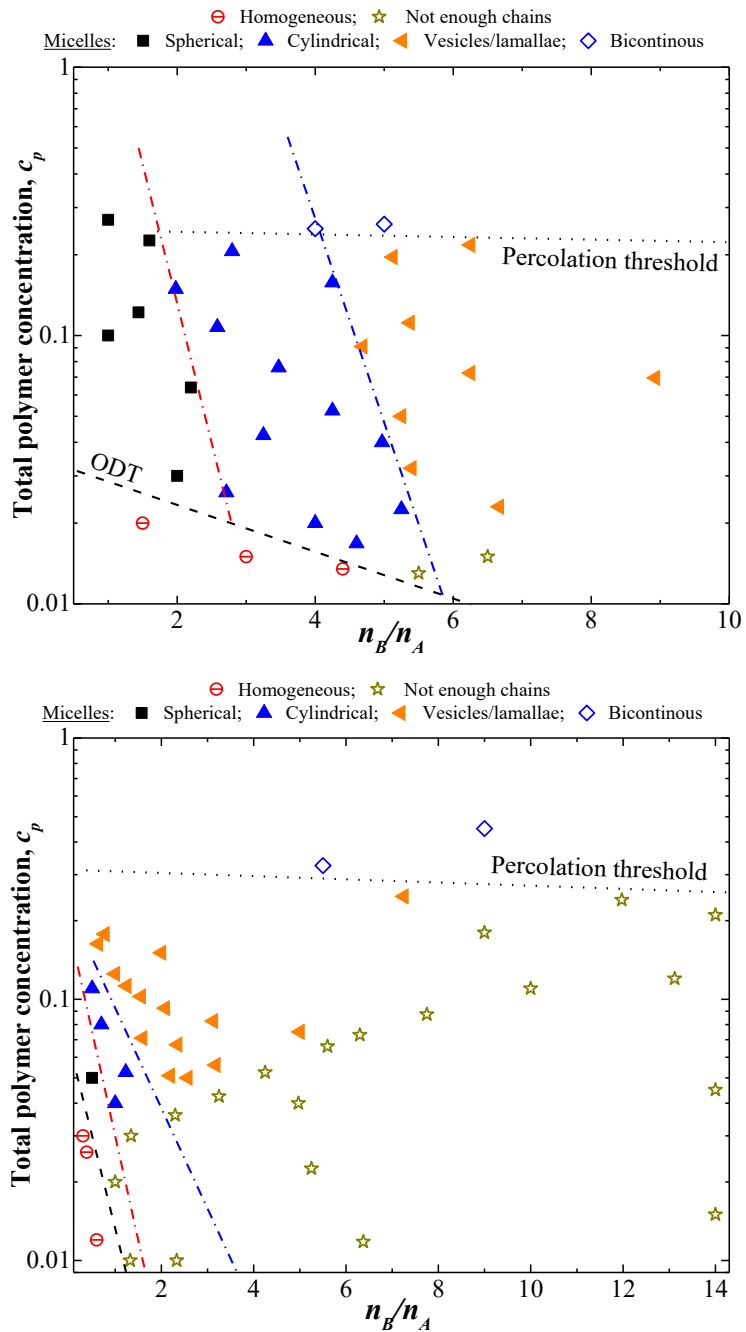

Fig.5 The phase diagram obtained for the PISA process with $p_i=0.1$ (top) and $p_i=0.01$ (bottom). The transition lines show the expected behavior and are depicted as a guide for the eye.

We see dramatic changes in the diagram: upon decreasing the initiation probability, the regions of spherical and cylindrical micelles shrink drastically, while the region of vesicles/lamellae expands, and at $p_i=0.01$ the majority of the obtained structures belong to that type. Let us now study the obtained polymer chains. Fig.6 shows the insoluble block length distributions for $c_p=5.25$ and $n_b/n_a=4.25$; this point was chosen so that different structures were observed at all three initiation probabilities: spherical micelles at $p_i=1$, cylindrical micelles at $p_i=0.1$ and vesicle at $p_i=0.01$.

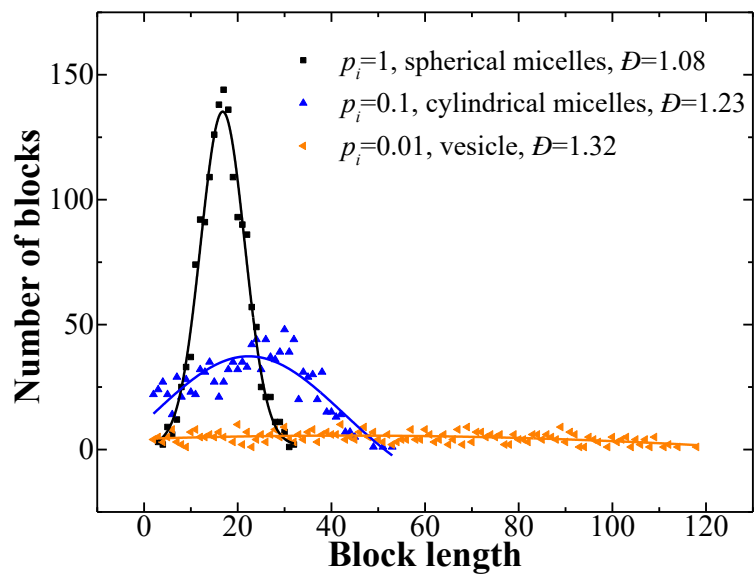

Fig.6 The insoluble block length distributions obtained at $c_p=5.25$ and $n_b/n_a=4.25$ obtained at three different values of $p_i$: 1)spherical micelles, $p_i=1$; 2) cylindrical micelles, $p_i=0.1$; 3) vesicle, $p_i=0.01$.

We can see that upon decreasing the initiation probability the distribution becomes noticeably wider, which is the result of the fact that the chains start growing at different moments. However, the calculated dispersity of 1.32 is not that high even at the slowest studied probability. It is also apparent form Fig.6 that the average soluble block length increases upon decreasing $p_i$. This is somewhat surprising given that the number of monomers per each precursor is fixed. A detailed analysis showed that the reason of the increase of the average block length is rather simple: not all the chains at low values of $p_i$ even have the insoluble block. In other words, since the initiation process is spread out in time at low values of $p_i$, some chains grow longer than other, and the chain that start to grow at the early stages of the process accumulate the most monomers. For example, for $p_i=0.01$ only 32% of the chain have an unsoluble block, while for $p_i=1$ all the chain have it. Therefore, decreasing $p_i$ essentially "remaps" the phase diagram coordinates by making the effective $c_p$ lower (by decreasing the number of active chains) and the effective $n_b/n_a$ higher (again, because the number of active chains gets higher, while the number of monomers remains the same). Of course, if we increase the length of precursor keeping the $n_b/n_a$ constant, the number of bare precursors will decrease as the distribution "moves" toward larger block lengths; one should keep in mind, however, that only chains with a long enough insoluble block tend to aggregate, and the others are dissolved. This means that not all the chains participate in the micelle formation, which, again, effectively increases $n_b/n_a$ and decreases $c_p$.

### 3.4 Percolating structures

As it was mentioned earlier, at sufficiently large polymer concentrations we observed percolation in the system (empty symbols in the diagrams). We can divide these structures into two general types: formed from cylinders (observed at intermediate values of $n_b/n_a$) and lamellae (observed at high values of $n_b/n_a$); typical snapshots of such systems are shown in Fig.2.

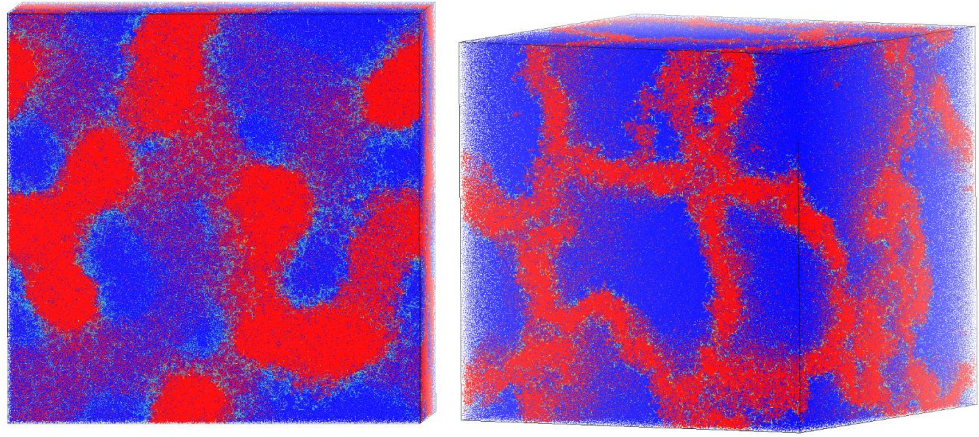

Fig.7. Typical snapshots of percolating systems formed from cylinders (left) and lamellae (right). The insoluble blocks are depicted in red, the soluble blocks are green, while the solvent is blue.

Such systems have not been previously described in the literature; they are of considerable interest due to the fact that they are essentially physical gels, which are formed due to the presence of large hydrophobic domains.

**3.5 Comparison to theory and experiments**

In the work [15] an analytical theory describing the micelle formation in solutions of diblock-copolymers was developed; the system is similar to that studied in Fig.1. Since we have shown that the ideal PISA process yields the same results (as seen in fig.1 and 3), we compare the results obtained during PISA with the aforementioned theory. Fig. 8 shows the position of the sphere-cylinder and cylinder-vesicle transitions obtained using the analytical theory [[15]] and our PISA simulations (Fig.3). The numerical coefficients of the analytical theory $C_f$ and $C_h$ were chosen so that the sphere-cylinder transition in theory and our simulations occurred at the same value of $n_b/n_a$.

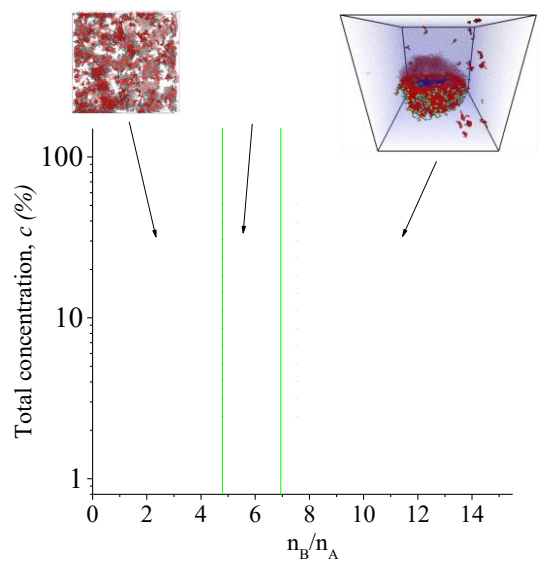

Fig.8 Comparison of the position of the sphere-cylinder and cylinder-vesicle transitions obtained using the analytical theory [15] and our PISA simulations.

We can see a rather good, but non-ideal, correspondence. The differences can be attributed to the finite chain lengths in simulations, due to which some of the assumptions made in the theory are not applicable. Another approach to compare our results to the predictions of the analytical theory is to define the combinations of $n_b$ and $n_a$ at which the sphere-cylinder and cylinder-vesicle transitions occur. In order to do that, we studied three additional values of $n_a$ – 2, 8 and 16. Fig. 9 shows the comparison of the theoretical scaling with the simulation data.

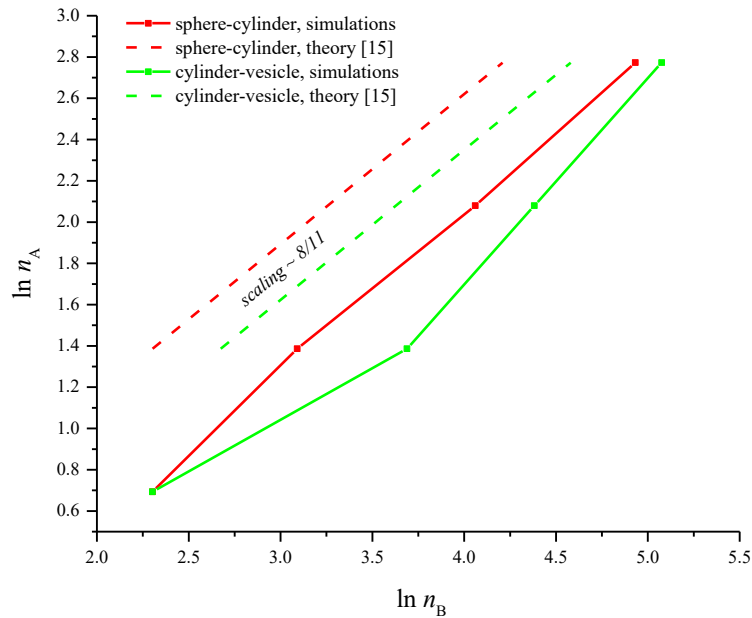

Fig.9 Comparison of the scaling of the sphere-cylinder and cylinder-vesicle transitions in terms of the block lengths obtained in the analytical theory [15] and our simulations.

Again, we see a very good correspondence.

As it was mentioned in the introduction, there is a number of experimental works on PISA. Since the main idea of this work is to shed some light on the behavior of the realistic systems, we think it is crucial to compare our results to the experimental realizations of PISA. In order to do that, we took the phase diagram obtained in [] and plotted it in the coordinates ($n_b/n_a$ - $c_p$). We found that the best correspondence between the experimental data and simulation results is obtained at $p_i$=0.1; the comparison is presented in Fig.10.

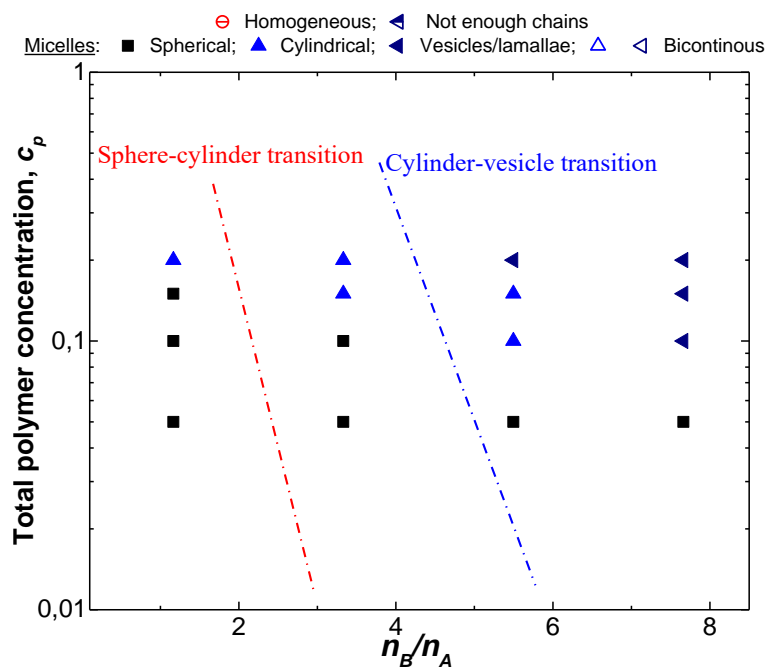

Fig.10. Comparison between the approximate transition lines obtained for PISA in simulations with $p_i$=0.1 (lines) and in experiments[] (points).

We can see that while the general trend is similar, there is some quantitative difference. However, comparing the phase diagram in Fig.3 with the experimental data in Fig.10, we can clearly see that taking into account the fact that the realistic reaction is not ideal allows us to achieve a better correspondence between simulations and experiments. Therefore, we can speculate that an even better correspondence could be achieved if the side reactions are allowed for; we plan to study it in our next work.

## 4. Conclusions

In this work we studied polymerization-induced self-assembly by means of computer simulations. First of all, a model based on the DPD method allowing variation of wide range of parameters was developed. Using this model, phase diagrams of the micelle states were constructed depending on the polymer concentration and the asymmetry of the composition for various reaction conditions. We found that if the reaction is ideal controlled radical polymerization (the initiation speed is much larger than the propagation speed and there are no side reactions such as termination or chain transfer), the phase diagram is no different from that obtained for pre-synthesized monodisperse diblock-copolymers with one insoluble block. Analysis of the insoluble block length distributions showed that the dispersity are very low (<=1.08), which indicates that the copolymers obtained during PISA are essentially monodisperse; that is obviously the reason why the diagrams coincide. Next, we studied two cases of slow initiation. We found that the phase diagram change dramatically: upon decreasing the initiation speed, the regions of spherical and cylindrical micelles shrink, while the region of vesicles/lamellae expands. This happens because at small initiation speed there is a significant amount of chains with a very short non-soluble block (and even without such block altogether), which do not participate in the formation of micelles. Therefore, decreasing the initiation speed essentially "remaps" the phase diagram coordinates by making the effective concentration lower

(by decreasing the number of active chains) and the effective block length ratio higher (again, because the number of active chains gets higher, while the number of monomers remains the same).

We found that when the polymer concentration is high enough, percolating structures emerge. Two types of such structures were observed: formed by cylindrical micelles or lamellae. Such structures are essentially a physical gel; they are promising candidates to be used as membranes for various applications.

The results were compared to the existing theoretical model of Zhulina and Borisov [15]. We found a good correspondence for the relative positions of the sphere-cylinder and cylinder-vesicle transitions as well as the scaling of the sphere-cylinder and cylinder-vesicle transitions in terms of the block lengths. Finally, we compared our results to one of the existing experimental phase diagrams; while the general trend is similar, there are some quantitative differences, which we attributed to the presence of side reactions in the real experiment.

Summarizing, in this work we shed some light on the very complex and interesting PISA process. We hope that our study will facilitate further investigation on this topic.